\documentclass[%
 reprint,
 superscriptaddress,   
 amsmath,amssymb,
 aps,prl
]{revtex4-2}

\makeatletter
\def\@makecaption#1#2{%
  \par
  \vskip\abovecaptionskip
  \begingroup
    \small\rmfamily
    \leftskip\z@ \rightskip\z@
    \parfillskip\z@ plus 1fil\relax  
    \noindent
    \@make@capt@title{#1}{#2}\par
  \endgroup
  \vskip\belowcaptionskip
}
\makeatother
\usepackage{xcolor}
\usepackage{dcolumn}
\usepackage{bm}
\usepackage{hyperref}
\usepackage{ragged2e}
\usepackage[justification=Justified, singlelinecheck=false]{caption}
\usepackage{subcaption}

\usepackage{float}
\usepackage{natbib}
\usepackage{ragged2e}

\usepackage[margin=1.5cm]{geometry}
\usepackage{tabularray}

\usepackage{graphicx}
\usepackage[tiny, center, uppercase]{titlesec}

\titlespacing\section{0pt}{12pt plus 0pt minus 0pt}{3pt plus 0pt minus 0pt}
\titlespacing\subsection{0pt}{12pt plus 4pt minus 2pt}{0pt plus 2pt minus 2pt}
\titlespacing\subsubsection{0pt}{12pt plus 4pt minus 2pt}{0pt plus 2pt minus 2pt}

\newcommand{\QMI}{\affiliation{Quantum Matter Institute, University of British Columbia, Vancouver, British Columbia, Canada V6T1Z4}}
\newcommand{\DPA}{\affiliation{Department of Physics \& Astronomy, University of British Columbia, Vancouver, British Columbia, Canada V6T1Z1}}
\newcommand{\MPI}{\affiliation{Max Planck Institute for Solid State Research, Heisenbergstr. 1, D-70569 Stuttgart, Germany}}
\newcommand{\KIT}{\affiliation{Institute for Quantum Materials and Technologies, Karlsruhe Institute of Technology, Kaiserstr. 12, 76131 Karlsruhe, Germany}}
\newcommand{\CAT}{\affiliation{Department of Physics and Institute for Quantum Information and Matter,
California Institute of Technology, Pasadena, California 91125, USA}}
\newcommand{\UCSB}{\affiliation{Materials Department, University of California, Santa Barbara, California 93106, USA}}
\newcommand{\CLS}{\affiliation{Canadian Light Source Inc., 44 Innovation Boulevard, Saskatoon, SK, S7N 2V3, Canada}}
\newcommand{\IOP}{\affiliation{Institute of Theoretical Physics, Chinese Academy of Sciences, Beijing 100190, China}}
\newcommand{\HNL}{\affiliation{Hefei National Laboratory, Hefei 230088, China}}
\newcommand{\UOT}{\affiliation{Department of Physics, University of Ottawa, Ottawa, ON, K1N 6N5, Canada}}
\usepackage[normalem]{ulem} 
\begin{document}
\preprint{APS/123-QED}
\title{Band-Selective Coupling to Charge Density Wave Fluctuations \\ in Kagome Superconductors A\texorpdfstring{V$_3$Sb$_5$}{V$_3$Sb$_5$}}
\setlength{\parskip}{0pt}
\setlength{\belowcaptionskip}{-10pt}

\author{P.C. Moen}\email{pmoen@phas.ubc.ca}          \QMI\DPA 
\author{T.G.H. Blank}      \QMI\DPA
\author{C. Le} \HNL
\author{S. Smit}           \QMI\DPA
\author{N. Heinsdorf}  \CAT
\author{S. Sarkar}         \QMI\DPA
\author{C.C. Au-Yeung}     \QMI\DPA
\author{V. Zimmermann}     \QMI\DPA\MPI
\author{J. Kraan}          \QMI\DPA
\author{C.S.B. Pang}     \QMI\DPA
\author{M. Tuniz}          \QMI\DPA
\author{M. Bluschke}       \QMI\UOT
\author{A.A. Haghighirad} \KIT
\author{I. Rajput} \KIT
\author{A.C. Salinas}      \UCSB
\author{S. Gorovikov}      \CLS
\author{P. Dosanjh}        \QMI\DPA
\author{J.I. Dadap}        \QMI\DPA
\author{A.K. Mills}        \QMI\DPA
\author{S. Zhdanovich}     \CLS
\author{M. Michiardi}      \QMI\DPA
\author{G. Levy}           \QMI\DPA
\author{I.S. Elfimov}        \QMI\DPA
\author{S.D. Wilson}       \UCSB
\author{M. Le Tacon} \KIT
\author{D.J. Jones}        \QMI\DPA
\author{X. Wu}         \IOP
\author{A. Damascelli}\email{damascelli@physics.ubc.ca}     \QMI\DPA

\date{\today}

\begin{abstract}
The kagome superconductors AV$_3$Sb$_5$ (A=K, Rb, Cs) host multiband superconductivity (SC) within an unconventional charge density wave (CDW) parent phase, yet the relationship between the two remains unresolved. Through time- and angle-resolved photoemission spectroscopy (tr-ARPES) experiments on CsV$_3$Sb$_5$ and KV$_3$Sb$_5$, we observe coherent band oscillations driven by coupling to a low-energy CDW amplitude mode at 3.1 THz. We find this coupling to be highly band-selective, differing by a factor of over 2 even between bands with dominant V-3d character. We reproduce these results with frozen-phonon density functional theory calculations, offering insight towards the relation between differing Fermi surface characters and our observed band selectivity. Furthermore, we observe differences in coupling channels between KV$_3$Sb$_5$ and CsV$_3$Sb$_5$. These results establish CDW amplitude mode fluctuations as a band-selective low-energy coupling channel that intimately contributes to the interplay between CDW and SC in the AV$_3$Sb$_5$ compounds. 
\end{abstract}

\maketitle

The kagome metals AV$_3$Sb$_5$ (A$=$K, Rb, Cs), with their V-based kagome lattice, host a complex phase diagram of intertwined electronic orders \cite{Wilson_Ortiz_2024}. A charge density wave (CDW) develops at T$_\mathrm{CDW} \,{\small \sim}\,$80–100~K, accompanied by signatures of time-reversal symmetry breaking whose microscopic origin remains debated \cite{Wilson_Ortiz_2024, Mielke2022, Suetsugu_2026, Xing2024, Jiang_2026}. At lower temperatures, the CDW coexists with multiband superconductivity (SC) below A-site dependent critical temperatures T$_c=$0.7–2.5~K, with the highest T$_c$ in CsV$_3$Sb$_5$ \cite{Wilson_Ortiz_2024, Chen_Chen_2022}, which also exhibits a nematic phase \cite{Li_2023, Guo_2025}. Reports of anisotropic, potentially nodal superconducting gaps on specific Fermi-surface sheets \cite{Almoalem_2026, Mine2025, Hossain_2025} further highlight the distinct roles of individual electronic bands. Although numerous theoretical proposals have explored how the CDW parent phase influences SC \cite{Holbaek_Fischer_2025, Jiang_Liu_Yu_2023, Jiang_Mao_Miao_Yu_Li_2024, Lin_Huang_Zhou_2024, Tazai_Yamakawa_Kontani_2026, Tazai_Yamakawa_Onari_Kontani_2022,Li_2025,Han_Zhan_Hu_Zhang_Wu_2026, Tian_Savrasov_2025, Schultz_Palle_Kim_Fernandes_Schmalian_2025, Mitra_Schultz_Kim_2025}, understanding the microscopic interplay between these orders and how it shapes their band-dependent properties remains a central challenge.

Electron–phonon coupling is a natural focus in addressing this challenge, as it can contribute both to charge ordering and to the attractive interaction responsible for superconducting pairing \cite{Zhong_2023, You_2025, Alkorta_2026, McGuinness_2026, Wang_2026}. In multiband systems such as the AV$_3$Sb$_5$ compounds, these contributions sensitively depend on which electronic bands couple to which lattice modes. The CDW itself can reshape this coupling by reconstructing the electronic bands and modifying the lattice excitation spectrum \cite{Luo2022, Lou2022, Liu2022, He2024}. Of particular interest are the CDW amplitude modes: collective oscillations of the magnitude of the coupled charge and lattice order. Their low energies and direct connection to the CDW make them a potentially important coupling channel for electrons in the superconducting phase. Establishing how these modes couple to individual bands is therefore essential towards assessing their role in the interplay between CDW order and multiband SC in the AV$_3$Sb$_5$ compounds.

\begin{figure*}[t]
\includegraphics[width=1.0\textwidth]{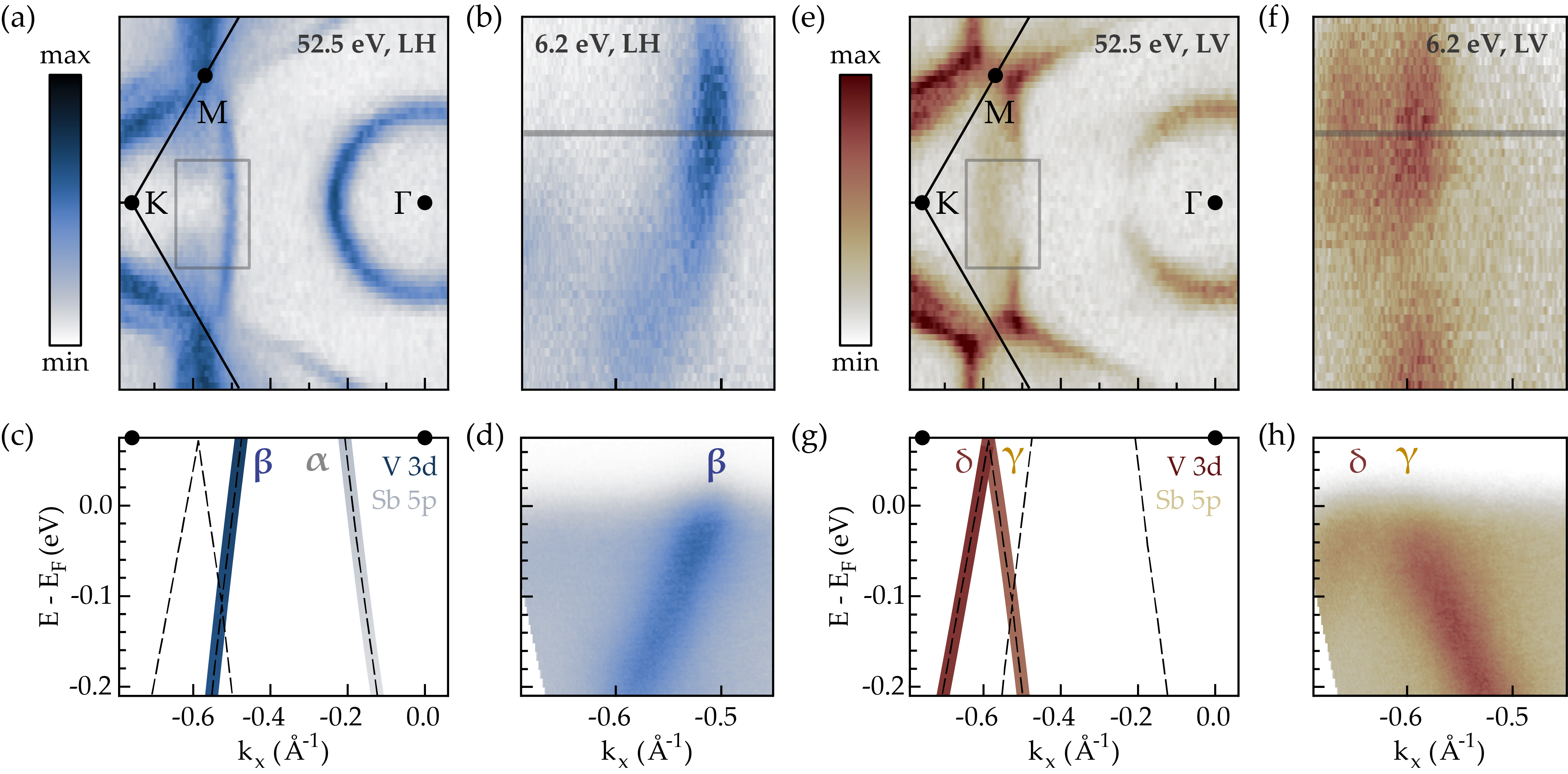}
\caption{(a) Fermi surface taken at $52.5$ eV, and (b) at $6.2$ eV spanning the boxed region in (a), with linear horizontal (LH) polarization. (c) Calculated band dispersion along the $\Gamma-K$ path; LH-visible bands are highlighted. (d) Energy-momentum cut along the grey line in (b). (e-h) Corresponding plots to (a-d) for linear vertical (LV) polarization. }\label{fig:fig1}
\phantomsubcaption\label{fig:fig1:a}
\phantomsubcaption\label{fig:fig1:b}
\phantomsubcaption\label{fig:fig1:c}
\phantomsubcaption\label{fig:fig1:d}
\phantomsubcaption\label{fig:fig1:e}
\phantomsubcaption\label{fig:fig1:f}
\phantomsubcaption\label{fig:fig1:g}
\phantomsubcaption\label{fig:fig1:h}
\end{figure*}

Several recent experimental and theoretical advances support this perspective. Inelastic x-ray scattering studies have established a central role for lattice dynamics in CDW formation \cite{McGuinness_2026, Wang_soft_2026}, while calculations and angle-resolved photoemission spectroscopy (ARPES) indicate electron–phonon coupling strengths sufficient to account for the superconducting transition temperature in CsV$_3$Sb$_5$ \cite{Zhong_2023, You_2025, Alkorta_2026}. Notably, the increase in T$_c$ in Nb-substituted samples is accompanied by a band-dependent enhancement of electron-phonon (e-ph) coupling \cite{Zhong_2023}. Raman spectroscopy provides complementary evidence for strong coupling between electronic excitations and CDW amplitude modes \cite{Liu2022, He2024}; one of these, an $A_{1g}$ mode near 103 cm$^{-1}$ (${\small\sim}$3.1~THz), softens and becomes strongly damped upon warming towards $\mathrm{T_{CDW}}$, with a Fano lineshape indicative of coupling to an electronic continuum. Time-resolved (tr) optical reflectivity experiments have coherently excited this mode and identified related collective excitations across the AV$_3$Sb$_5$ family \cite{Ratcliff2021, Deng_2025, Deng_K2025}, while tr-ARPES has observed amplitude modes in the CDW gap dynamics \cite{Azoury_2023, Liu_2025}. Together, these findings highlight CDW amplitude modes as a specific low-energy coupling channel whose contribution must be resolved at the level of individual electronic bands.

Despite these advances, a band- and mode-resolved study of the amplitude-mode contribution to e-ph coupling remains lacking. Here we address this by combining the orbital sensitivity of ARPES linear dichroism \cite{Hu_Wu2022} with the ability of tr-ARPES to track the electronic response to coherently excited collective modes \cite{Boschini_2024}. This approach isolates individual bands and tracks their energy modulation by a common collective displacement; under identical excitation conditions, the amplitude ratios of these modulations reveal differences in mode-specific deformation potentials and hence band-selective e-ph coupling \cite{Gerber_2017}. In CsV$_3$Sb$_5$, we identify coherent oscillations driven by a 3.1~THz CDW amplitude mode, with inferred coupling varying by over a factor of 2 between even predominantly V-3d bands. Frozen-phonon calculations reproduce these variations in coupling strength, linking the selectivity to the response of individual bands to the amplitude-mode lattice distortion. In KV$_3$Sb$_5$, we observe an additional band-selective response near 3.5~THz, absent in CsV$_3$Sb$_5$, revealing distinct low-energy coupling channels in nominally similar compounds. These findings highlight the importance of resolving CDW collective dynamics at the level of individual bands and modes, and constrain how amplitude fluctuations could influence pairing on different Fermi-surface sheets in the AV$_3$Sb$_5$ superconductors.

\begin{figure*}[t]
\centering
\captionsetup{width=\textwidth}
\includegraphics[width=\textwidth]{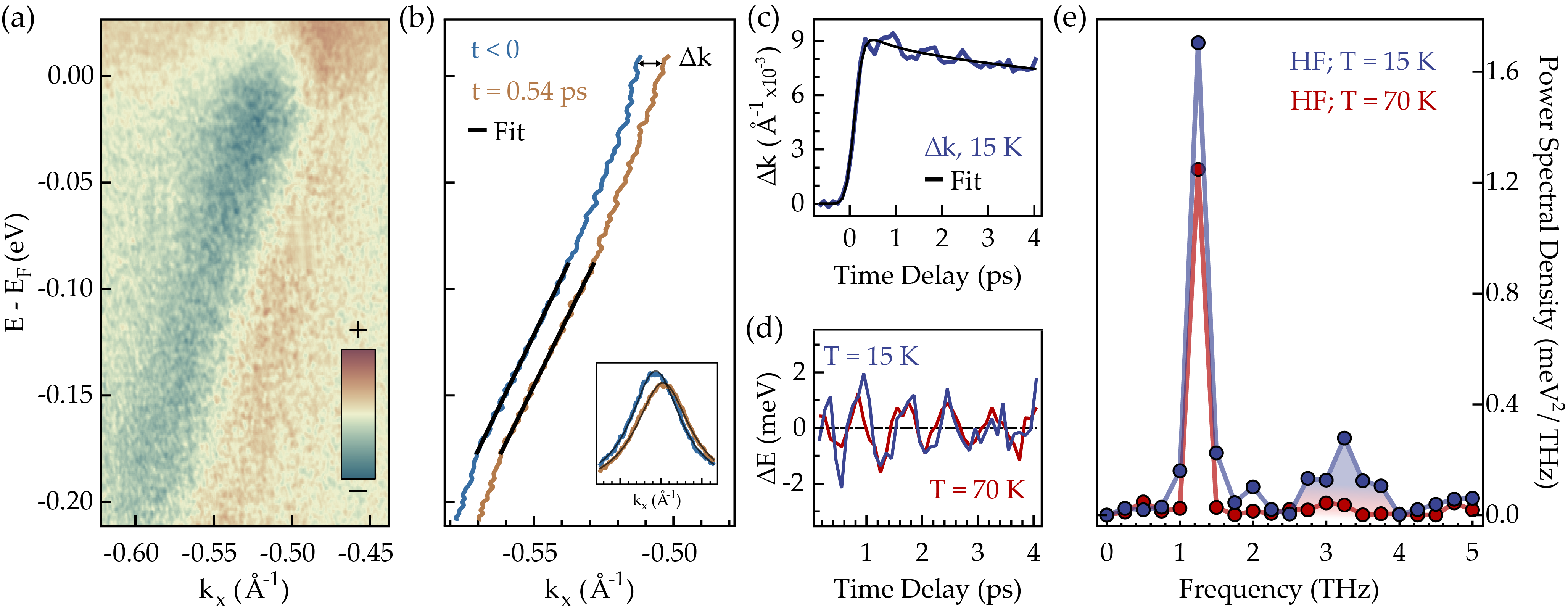}
\caption{(a) Intensity difference plot of the $\beta$ band, $I(t\!=\!0.54 \text{ ps})\!-\!I(t\!<\!0)$, each binned $300 \text{ fs}$. (b) Energy-momentum dispersion extracted from peak positions of Lorentzian MDC fits (inset). (c) Change in momentum position of band averaged over linearly fitted area of dispersion in (b), with phenomenological rise-and-decay fit. (d) Residuals of fits to band shifts for low and high base temperature, showing coherent oscillations; these are converted to energy by multiplying by the band velocity obtained from the linearly fitted region in (b). (e) Power spectral density (PSD) from Fourier transform of (d), showing mode-resolved contributions to coherent oscillations for T$\!=\!15$ K and $70$ K base temperature.}\label{fig:fig2}
\phantomsubcaption\label{fig:fig2:a}
\phantomsubcaption\label{fig:fig2:b}
\phantomsubcaption\label{fig:fig2:c}
\phantomsubcaption\label{fig:fig2:d}
\phantomsubcaption\label{fig:fig2:e}
\end{figure*}
Single crystals of CsV$_3$Sb$_5$ and KV$_3$Sb$_5$ were obtained from multiple sources, as described elsewhere \cite{Ortiz_Teicher_2021, Stier_2024}. Equilibrium ARPES measurements were conducted at the Quantum Materials Spectroscopy Centre (QMSC) beamline at the Canadian Light Source with a Scienta-Omicron R4000 hemispherical analyzer, at a base temperature of 10 K and with an energy resolution better than 20 meV. Tr-ARPES measurements were conducted at the UBC–Moore Center for Ultrafast Quantum Matter using a Scienta-Omicron DA30 analyzer, using an LH/LV-polarized 6.2 eV probe (where LH/LV refer to linear horizontal/vertical, respectively) and an LV-polarized 1.55 eV pump with a FWHM of 210 fs, with total energy and time resolutions better than 40 meV and 300 fs \cite{Dufresne_2024}. High- and low-fluence (HF/LF) measurements were performed at probe-weighted fluences of 365 and 161 $\mathrm{\mu J/cm^2}$, respectively.

To start, we characterize the band structure of CsV$_3$Sb$_5$ by performing equilibrium ARPES measurements with emphasis along the $\Gamma$-K direction in the Brillouin zone. As shown in \hyperref[fig:fig1]{Fig.~\ref*{fig:fig1}},  there is a high degree of matrix element selectivity along this high-symmetry direction: LH polarization emphasizes the hexagonal and zone-center Fermi-surface pockets, while suppressing the two triangular zone-edge pockets [\hyperref[fig:fig1:a]{Fig.~\ref*{fig:fig1}(\subref*{fig:fig1:a})}]; comparison with our density functional theory (DFT) calculations identifies the corresponding $\beta$ and $\alpha$ bands as predominantly V-3d and Sb-5p, respectively [\hyperref[fig:fig1:c]{Fig.~\ref*{fig:fig1}(\subref*{fig:fig1:c})}; the zero of energy is at the Fermi energy]. LV polarization instead enhances the triangular pockets [\hyperref[fig:fig1:e]{Fig.~\ref*{fig:fig1}(\subref*{fig:fig1:e})}], associated with the predominantly V-3d-derived $\delta$ and $\gamma$ bands [\hyperref[fig:fig1:a]{Fig.~\ref*{fig:fig1}(\subref*{fig:fig1:g})}]. This selectivity persists at the 6.2 eV probe energy used in our tr-ARPES experiments, as confirmed by both the Fermi-surface maps and $\Gamma$-K dispersions in \hyperref[fig:fig1:b]{Fig.~\ref*{fig:fig1}(\subref*{fig:fig1:b},\subref*{fig:fig1:d},\subref*{fig:fig1:f},\subref*{fig:fig1:h})}, enabling measurements of individual band responses. Because the $\delta$ band has much lower photoemission intensity than $\alpha$, $\beta$ and $\gamma$, we restrict our analysis to the latter bands.

Turning to the time domain, we first focus on the $\beta$ band. \hyperref[fig:fig2:a]{Figure~\ref*{fig:fig2}(\subref*{fig:fig2:a})} displays the difference in intensity between $t=0.54$ ps after the pump pulse and before $t=0$ ps, revealing an apparent pump-induced shift towards lower energy, or equivalently smaller $\mathrm{|k_x|}$, seen as an intensity loss along the equilibrium dispersion and a gain alongside it [\hyperref[fig:fig2:a]{Fig.~\ref*{fig:fig2}(\subref*{fig:fig2:a})}]. To quantify this shift, we fit the momentum distribution curves (MDCs) to Lorentzian lineshapes over a broad energy range [\hyperref[fig:fig2:b]{Fig.~\ref*{fig:fig2}(\subref*{fig:fig2:b})}, inset], at all measured delays; the fitted peak positions yield the dispersions before and after excitation, as shown in \hyperref[fig:fig2:b]{Fig.~\ref*{fig:fig2}(\subref*{fig:fig2:b})}. We then track the momentum shift $\Delta k(t)$ by averaging over the linearly fitted region in \hyperref[fig:fig2:b]{Fig.~\ref*{fig:fig2}(\subref*{fig:fig2:b})}, away from the Fermi level to avoid energy-resolution effects \cite{Levy2014}; the resulting time dependence shows a rapid pump-induced shift of $0.009 \: \mathrm{\AA^{-1}}$, followed by slower relaxation [\hyperref[fig:fig2:c]{Fig.~\ref*{fig:fig2}(\subref*{fig:fig2:c})}]. A phenomenological rise-and-decay model [see Supplemental Material (SM)] captures these dynamics [black line, \hyperref[fig:fig2:c]{Fig.~\ref*{fig:fig2}(\subref*{fig:fig2:c})}]. Subtracting the fit reveals oscillations on top of the rigid band shift [\hyperref[fig:fig2:d]{Fig.~\ref*{fig:fig2}(\subref*{fig:fig2:d})}], where we convert the residual momentum shifts to energy shifts by multiplying by the slope of the linear dispersion fit (i.e., the band velocity) from \hyperref[fig:fig2:b]{Fig.~\ref*{fig:fig2}(\subref*{fig:fig2:b})}.

\begin{figure*}[t]
\centering
\captionsetup{width=1\textwidth}
\includegraphics[width=\textwidth]{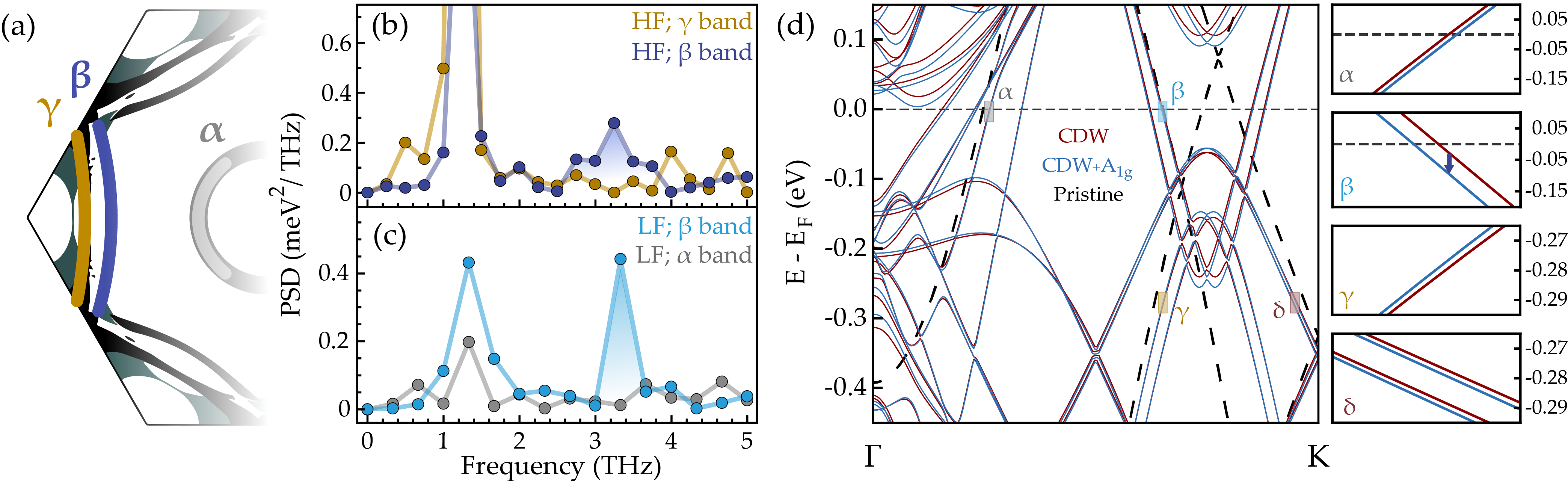}
\caption{(a) Calculated Fermi surface for the pristine phase, with the $\gamma$, $\beta$, and $\alpha$ Fermi surface sheets outlined. (b) High (HF, 365 $\mu$J/cm$^2$) and (c) low (LF, 161 $\mu$J/cm$^2$) pump fluence PSD, comparing $\gamma/\beta$ and $\alpha/\beta$, respectively. (d) Band structure calculations for the pristine and CDW phases, as well as CDW with a frozen 3.1~THz $\mathrm{A_{1g}}$ amplitude mode distortion (CDW+$\mathrm{A_{1g}}$); the resulting band shifts in the highlighted regions are shown in the panels to the right.}\label{fig:fig3}
\phantomsubcaption\label{fig:fig3:a}
\phantomsubcaption\label{fig:fig3:b}
\phantomsubcaption\label{fig:fig3:c}
\phantomsubcaption\label{fig:fig3:d}
\end{figure*}

To gain quantitative insight into these coherent oscillations, we plot the power spectral density (PSD) in \hyperref[fig:fig2:e]{Fig.~\ref*{fig:fig2}(\subref*{fig:fig2:e})}. The 15~K spectrum [\hyperref[fig:fig2:e]{Fig.~\ref*{fig:fig2}(\subref*{fig:fig2:e})}, blue] is dominated by peaks near 1.3 and 3.1~THz: while the ${\small\sim}1.3$~THz mode corresponds to a folded $Q\!=\!L$ phonon involving out-of-plane Cs vibrations, observed by tr-optical-reflectivity \cite{Ratcliff2021, Deng_2025}, the 3.1~THz mode matches the CDW amplitude mode observed by Raman spectroscopy near 103~cm$^{\small -1}$ \cite{Deng_2025}. To test this assignment, we repeat the measurements at $\mathrm{T}\!=\!70$~K, closer to $T_{\mathrm{CDW}}$, where strong amplitude-mode damping is expected \cite{Liu2022, He2024, Deng_2025}. The same analysis shows that the 1.3~THz folded phonon persists, while the 3.1~THz response becomes almost undetectable [\hyperref[fig:fig2:d]{Fig.~\ref*{fig:fig2}(\subref*{fig:fig2:d},\subref*{fig:fig2:e})}, red]. This pronounced temperature dependence is atypical of a conventional phonon, and agrees with Raman \cite{He2024} and tr-optical-reflectivity measurements \cite{Ratcliff2021, Deng_2025}, supporting the identification of the 3.1~THz feature as a CDW amplitude mode.

We next extend the analysis to the $\gamma$ and $\alpha$ bands [\hyperref[fig:fig3:a]{Fig.~\ref*{fig:fig3}(\subref*{fig:fig3:a})}]. Surprisingly, contrary to the $\beta$ band, the $\gamma$ band shows no resolvable peak near 3.1~THz [\hyperref[fig:fig3:b]{Fig.~\ref*{fig:fig3}(\subref*{fig:fig3:b})}]. Integrating the PSD over 2.7--3.5~THz and taking the square root gives a root sum squared (RSS) oscillation magnitude over 2 times larger for the $\beta$ band compared to the $\gamma$ band ($2.3\pm 0.5$), despite their shared dominant V-3d character. Our low-fluence measurements likewise reveal a stronger amplitude-mode response for the $\beta$ than the $\alpha$ band [\hyperref[fig:fig3:c]{Fig.~\ref*{fig:fig3}(\subref*{fig:fig3:c})}], with an over 3 times larger RSS oscillation magnitude ($3.2\pm 0.2$) in the 3.0--3.5~THz range.

To confirm the observed band selectivity, we perform frozen-phonon DFT calculations on CsV$_3$Sb$_5$ in the CDW phase (see SM for details). We first identify the 3.1~THz $A_{1g}$ mode in our $Q\!=\!0$ phonon spectrum as a CDW amplitude mode by matching its atomic displacement pattern to that reported in Ref.~\onlinecite{Liu2022}. We then compare the equilibrium CDW band structure with that obtained by imposing this mode as a frozen distortion on the CDW structure [\hyperref[fig:fig3:d]{Fig.~\ref*{fig:fig3}(\subref*{fig:fig3:d})}]. The resulting energy shifts in the calculated dispersions quantify the relative coupling strengths to various bands: these are displayed in the right-hand-side panels of \hyperref[fig:fig3:d]{Fig.~\ref*{fig:fig3}(\subref*{fig:fig3:d})} for regions where we can identify the calculated bands with those observed by our ARPES measurements. Evidently, the $\beta$ band has the largest response, in agreement with our experimentally observed band selectivity. In particular, the $\beta$ band experiences an energy shift larger than the $\delta$, $\gamma$, and $\alpha$ bands by ratios of 3.38, 2.56, and 3.18, respectively, supporting our identification of the 3.1~THz $A_{1g}$ CDW amplitude mode as the band-selective low-energy coupling channel observed in our tr-ARPES measurements.

The microscopic implications of these results can be understood by considering the deformation potential $D_{n\nu}\!=\!\partial E_n/\partial Q_\nu$, where $E_n$ is the energy of band $n$ and $Q_\nu$ is the displacement coordinate of the $A_{1g}$ phonon mode $\nu$ \cite{Gerber_2017}. Under identical excitation conditions, all bands respond to the same coherent lattice motion $Q_\nu(t)$ regardless of probe polarization. Since $\delta E_n(t)\!=\!D_{n\nu}\times\delta Q_\nu(t)$, differences in the amplitude of band oscillations must then correspond to differences in the magnitude of the deformation potential. In turn, for a given mode, the deformation potential is proportional to the corresponding e-ph coupling matrix element $g_\nu$ \cite{Bardeen_Shockley_1950}. Thus, our experiment directly accesses the ratio of $g_\nu$ between different bands, constituting a microscopic measure of band-selective e-ph coupling.


Since the $A_{1g}$ amplitude mode we observe strongly modulates $2\times2$ charge-bond order \cite{Liu2022}, its band selectivity can be understood from theoretical studies of the ideal nesting conditions for this kagome lattice instability~\cite{WuPRL2021,FuNSR2025,ZhanPRL2026}. Among the kagome-derived bands in CsV$_3$Sb$_5$, the hexagonal Fermi surfaces of $\beta$ and $\gamma$ offer more favourable nesting conditions than that of $\delta$, which is warped by a higher-order van Hove singularity (vHS). However, sublattice-dependent scattering matrix elements favour nesting specifically on the $\beta$ band. This preference arises from sublattice effects intrinsic to the kagome lattice, where different types of vHS are associated with distinct sublattice characters. For a filling close to a p-type vHS, the Fermi surface acquires an alternating sublattice character that favours such bond modulations~\cite{FuNSR2025,ZhanPRL2026}. The dual-sublattice character of $\beta$ along $\Gamma{\small -}$K (see SM) is consistent with this filling, whereas the three-sublattice character of $\gamma$ points to an m-type vHS filling. These considerations single out the $\beta$ band as the preferred coupling channel for fluctuations of the $2\times2$ CDW order in the AV$_3$Sb$_5$ compounds.

\begin{figure}[b]
\captionsetup{width=\linewidth}
\includegraphics[width=\linewidth]{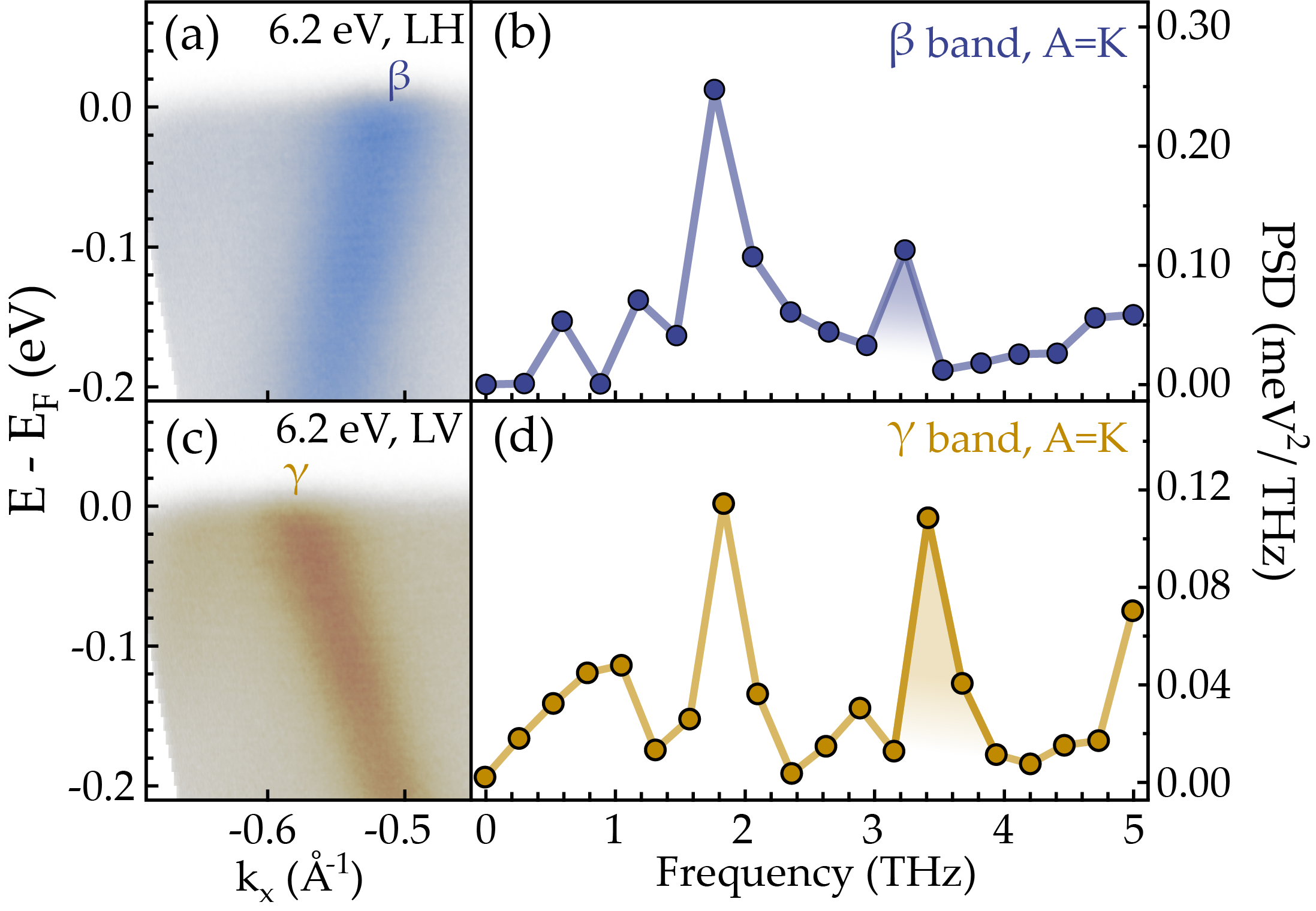}
\caption{(a) $\beta$ band dispersion of KV$_3$Sb$_5$ and (b) corresponding mode-resolved coherent oscillation analysis, obtained with LH-polarized 6.2 eV photons. (c,d) Same as (a,b) for the $\gamma$ band, taken with LV polarization.}\label{fig:fig4}
\phantomsubcaption\label{fig:fig4:a}
\phantomsubcaption\label{fig:fig4:b}
\phantomsubcaption\label{fig:fig4:c}
\phantomsubcaption\label{fig:fig4:d}
\end{figure}
For comparison between AV$_3$Sb$_5$ variants, we perform similar experiments on KV$_3$Sb$_5$, which exhibits a lower T$_c\!=\!0.9$ K compared to CsV$_3$Sb$_5$ as well as a distinct superconducting phase diagram \cite{Wilson_Ortiz_2024}. As before, we utilize different probe polarizations to isolate the $\beta$ and $\gamma$ bands [\hyperref[fig:fig4:a]{Fig.~\ref*{fig:fig4}(\subref*{fig:fig4:a},\subref*{fig:fig4:c})}]. Their coherent oscillation responses [\hyperref[fig:fig4:b]{Fig.~\ref*{fig:fig4}(\subref*{fig:fig4:b},\subref*{fig:fig4:d})}] include unique features compared to those of CsV$_3$Sb$_5$: ($i$) The folded phonon shifts from ${\small\sim}$1.3 to ${\small\sim}$2~THz owing to the lighter K ions \cite{Deng_K2025}. ($ii$) The $\beta$ band retains a band-specific response near 3.1~THz, consistent with the CDW amplitude mode identified by Raman spectroscopy and tr-optical-reflectivity \cite{Oh_2026, Deng_K2025}. ($iii$) An additional response -- completely absent in CsV$_3$Sb$_5$ -- appears near 3.5~THz only on the $\gamma$ band; this feature has a reported anomalous temperature dependence \cite{Deng_K2025} and matches a calculated CDW amplitude-mode frequency for KV$_3$Sb$_5$ (see SM). Given the broad similarities in the electronic structure and CDW distortions of the AV$_3$Sb$_5$ compounds \cite{Lan2026, Kautzsch_2023}, this difference in coupling channels is striking. A closer inspection of our calculated CDW band structures, however, reveals a marked difference between the K and Cs compounds (see SM). In CsV$_3$Sb$_5$, the CDW gap along $\Gamma$--K reconstructs the $\gamma$ band to below $E_F$ more strongly than the $\delta$ band, leaving the latter with a larger low-energy contribution. This hierarchy reverses in KV$_3$Sb$_5$: the $\gamma$ band retains its Fermi-level crossing because its CDW gap lies entirely above $E_F$, whereas the $\delta$ band crossing is strongly gapped. These contrasting reconstructions highlight how CDW order can reshape the low-energy band and orbital character of nominally similar compounds.

Altogether, our results establish that e-ph coupling to CDW amplitude modes is highly band-selective in the AV$_3$Sb$_5$ kagome superconductors. A large $g_{kk'}^\nu$ is only one factor governing a mode's contribution to the total e-ph coupling constant. Nevertheless, the strong response of the $\beta$ band to the 3.1~THz amplitude mode is consistent with the strong coupling to a ${\small\sim}13$~meV (${\small \sim}$3.1 THz) mode observed in the Eliashberg function of the $\beta$ band as obtained from high-resolution ARPES measurements \cite{Zhong_2023}. Nb substitution also enhances the overall coupling strength of the $\beta$ band, largely through an increase in the Eliashberg function around this energy \cite{Zhong_2023}. This correspondence points to CDW amplitude modes as significant microscopic contributors to the band-dependent low-energy physics of these compounds, underscoring their importance towards the emergence of SC.

This band selectivity is also interesting in the context of possible loop-current-mediated SC, as different CDW amplitude modes could be intertwined with distinct patterns of loop-current fluctuations and thereby influence the superconducting order parameter \cite{Schultz_Palle_Kim_Fernandes_Schmalian_2025, Mitra_Schultz_Kim_2025}. CDW amplitude modes could also play a role in the recently proposed scenario of frustrated local moments \cite{Jiang_2026}: their associated distortions of the kagome lattice may relieve frustration, offering a possible structural analogue to the effects of Sn substitution. Such effects could remain relevant beyond the ordered CDW phase, given evidence that CDW fluctuations persist at doping levels far beyond the suppression of long-range order \cite{Kongruengkit_2026}.

Alongside these possibilities, the material-dependent coupling to these modes warrants further study, as it may have important consequences for the band dependence of SC. Our results emphasize the importance of band-dependent behaviour in the AV$_3$Sb$_5$ materials family, even between kagome-derived bands. In a broader context, our identification of low-energy coupling to CDW amplitude modes suggests that the relationship between CDW and SC can extend beyond mere competition. Although a competition between these orders is present in the AV$_3$Sb$_5$ compounds \cite{Alkorta_2026}, our results demonstrate that the CDW also introduces band-selective amplitude-mode coupling channels, opening a route through which its collective dynamics could contribute to superconducting pairing.
\\

\textit{Acknowledgements}---This research was undertaken thanks in part to funding from the Max Planck-UBC-UTokyo Centre for Quantum Materials and the Canada First Research Excellence Fund, Quantum Materials and Future Technologies. This project is also funded by the Natural Sciences and Engineering Research Council of Canada (NSERC), Canada Foundation for Innovation (CFI); the Department of National Defence (DND); the British Columbia Knowledge Development Fund (BCKDF); the Mitacs Accelerate Program; the QuantEmX Program of the Institute for Complex Adaptive Matter; the Gordon and Betty Moore Foundation's EPiQS Initiative, Grant GBMF4779 to A.D. and D.J.J.; the Canada Research Chairs Program (A.D.); the CIFAR Quantum Materials Program (A.D.); and the John Simon Guggenheim Memorial Foundation (A.D.). Use of the Canadian Light Source (Quantum Materials Spectroscopy Centre), a national research facility of the University of Saskatchewan, is  supported by CFI, NSERC, the National Research Council, the Canadian Institutes of Health Research, the Government of Saskatchewan and the University of Saskatchewan. T.G.H.B. acknowledges support by Rubicon 2024-2 with file number 019.242EN.003, which is financed by the Dutch Research Council (NWO) under the grant https://doi.org/10.61686/SGRRI73808. S.D.W. and A.C.S. acknowledge support via the UC Santa Barbara NSF Quantum Foundry funded via the Q-AMASE-i program under Award No. DMR-1906325. A.A.H. and M.L.T. acknowledge funding from the Deutsche Forschungsgemeinschaft (DFG, German Research Foundation), grant no. TRR288–422213477 (project B03). X.W. is supported by the National Key R\&D Program
of China (Grant No. 2023YFA1407300) and the National
Natural Science Foundation of China (Grant Nos. 12574151,
12447103, and 12447101).
\bibliographystyle{apsrev4-2}
\bibliography{apssamp}

\end{document}